# BEAM DYNAMICS STUDIES FOR THE CLIC MAIN LINAC

I. Nesmiyan, R.M. Jones,
School of Physics and Astronomy, The University of Manchester, Manchester, U.K
The Cockcroft Institute of Accelerator Science and Technology, Daresbury, U.K
A. Latina, D. Schulte, CERN, Geneva, Switzerland

*Abstract*

The implications of long-range wakefields on the beam quality are investigated through a detailed beam dynamics study. Injection offsets are considered and the resulting emittance dilution recorded, including systematic sources of error. These simulations have been conducted for damped and detuned structures (DDS) and for waveguide damped structures-both for the CLIC collider.

## INTRODUCTION

The Compact Linear Collider (CLIC) [1] is designed for electron-positron collisions at a 3 TeV centre of mass energy in the baseline design – although there also exist a preliminary design for 0.5 TeV collisions. CLIC relies on a two-beam concept in which a high current drive beam is decelerated and serves as an rf field for the main accelerated beam – in this way the number of klystrons needed are substantially reduced. Essentially this can be viewed as a transformer in which the 100 A drive beam is transformed into the 1 A accelerated beam. The accelerated beam consists of 312 bunches spaced from their immediate neighbours by 0.5 ns – each bunch of each is populated with ~$3.7 \times 10^9$ particles.

The head of each bunch in the train excites a wakefield $W_t$, which in principle consist of an infinite series of eigenmodes. This wakefield has both short-range and long-range components. The short-range wakefield acts over the bunch itself and in this case $W_t \propto <a>^{-3}$ [2], where $<a>$ is average iris radius. Once the geometrical parameters have been designed the short-range wakefield is then fixed. However, the long-range wakefield, which affects neighbouring bunches, can be suppressed. There are several methods by which this can be achieved. CLIC_G [2,3] is the baseline design, which relies on heavy damping through waveguides attached to each accelerating cell (Q ~10).

Two main alternative designs are also being investigated at present: choke-mode damping [4], and Damped Detuned structure (DDS) [5].

The choke-mode scheme relies on all of the higher order modes flowing out through essentially radial waveguides and the accelerating mode is reflected back into the structure. Wakefield suppression is in the progress of being optimized [4]. The DDS scheme relies on strong detuning of the dipole modes together with moderate damping (Q ~500-1500) affected by four waveguide-like manifolds, which run parallel to the acceleration axis of the beam. An additional feature of the DDS is that by monitoring the energy radiated to the manifolds, both the beam position and the cell-to-cell alignment can be remotely determined [6]. However, in order to properly suppress the wakefield many cells are needed to sample the Gaussian distribution in frequency space. As we rely on a similar number of cells as the CLIC baseline design this necessitates interleaving of the modes of neighbouring structures. It is worth emphasising that all of these designs must minimize the surface e.m. field in order to ensure electric breakdown does not occur.

In order to assess the impact of these beam-exited wakefields on the beam quality we conducted beam dynamics simulations with the code PLACET [7]. We have also utilized an approximate analytical formalism [8], to rapidly obtain a number of figures of merit for structure based on beam dynamics. In the next section the method is outlined, followed by a section on detailed beam dynamics studies for a series of DDS geometries.

## ANALYTICAL MODEL OF BEAM DYNAMICS

In order to assess the beam quality after its progress throughout the complete ~21 km linac we use the code PLACET. In addition, in order to rapidly analyse the effect of the wakefield of the beam under various conditions, we utilize an analytical method with several simplifying assumptions [8]. The model is based on point-like bunches progressing through a lattice while overall effect will be averaged. The beta function is assumed to depend on energy as $\beta \propto E^{1/2}$.

The analysis starts by considering bunch $k$ with the initial offset $y_k(0)$ which kicks bunch $j$. The final offset of bunch $j$ is:

$$y_j(L) = (1 + a_{jk}) y_k(0),$$

where

$$a_{jk} = \begin{cases} i\int_0^L \frac{W(z_j - z_k) N_e e^2 \beta(s)}{2E(s)} ds = iW(z_j - z_k) N_e e^2 \left\langle \frac{\beta}{E} \right\rangle_{eff}, & \forall j > k \\ 0, & \forall j \leq k \end{cases}$$

L is the length of the linac, $N_e$ is the number of $e^-$ in the bunch, $W(z_j - z_k)$ is transverse wakefield exited by bunch $k$ and experienced by bunch $j$. This analysis describes the *direct impact* of the bunch on another. However, to include the influence of succeeding bunches on one other – which we refer to as the *indirect effect*:

$$y_j(L) = \lim_{n_p \to \infty} (1 + a_{jk}/n_p)^{n_p} y_k(0) = \exp(a_{jk}) y_k(0) = A_{jk} y_k(0)$$

Matrix A includes both the *direct* and *indirect* effect. To study the impact of the long-range wakefield on the beam we use the following variables: $F_c$ which describes *coherent jitter*, $F_{rms}$ which assesses *random bunch-to-bunch jitter*, where

$$F_c = \frac{y_f y_f^*}{y_i y_i^*} = \frac{1}{n_p}\sum_k \left|\sum_j A_{kj}\right|^2, \quad F_{rms} = \frac{1}{n_p}\sum_{k=0}^{n_p-1}\sum_{j=1}^{k}|A_{jk}|^2.$$

A comparison is shown in Fig. 1 for the CLIC_G baseline design. The agreement between the simulations and the analytical method is excellent. This provides some validation of the analytical approach.

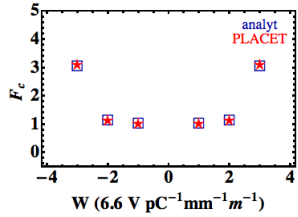

Figure 1: $F_c$ as a function of the wakefield at the first trailing bunch obtained analytically and by using the code PLACET for the point-like bunches at the end of the CLIC main linac for an initial 2σ offset in the bunch train.

In practice provided the first trailing bunch is below $6.6V(pC\cdot m\cdot mm)^{-1}$, $F_c \approx 1$ and $F_{rms} \approx 5$, then the emittance dilution is kept within acceptable bounds.

In the next section we apply this analytical technique to provide guidance on the wakefield suppression needed in the CLIC DDS structures.

## BEAM DYNAMICS ANALYSIS

Gaussian detuning of the modal frequencies allows the wakefield, which in the short-range is the inverse Fourier transform of the kick factor weighted density function, to fall in a Gaussian manner. Sampling the frequency distribution with an infinite number of cells ensures a Gaussian fall-off in the wakefield. However, in practice a finite number of cells sample the frequency distribution. Consequently the modes, which constitute the wakefield, will recohere at some point. This recoherence position is proportional to the number of cells [9]. Clearly it is advantageous to increase the point of recoherence. This is achieved by interleaving the modes of the successive structures. We fix the number of cells to those of the CLIC_G baseline design, namely 24, and interleaved these with 7 additional structures to move the re-coherence point from 2 m to 16 m (illustrated in Fig. 2). However, as the bunch train is 46.8 m long we require additional damping. This is facilitated through waveguide-like manifolds coupled through slots to each cell. The wakefield in the present 8-fold interleaved design, shown in the Fig. 2(b), is not adequately suppressed, as the indirect effect is particularly severe in this case. However, if we increase the effective wakefield damping, by imposing a Q~700, then the wakefield, illustrated in Fig. 3(a) is sufficient to satisfy the beam dynamics requirements – illustrated in Fig. 3(b). The bandwidth of these structure, $\Delta\omega/2\pi$ =2.1GHz=3.48σ$_g$, is the result of an optimization from structures of various bandwidths.

In fabricating several thousands of these accelerating structures there will inevitably be both systematic and random errors, which will occur due to the machining process.

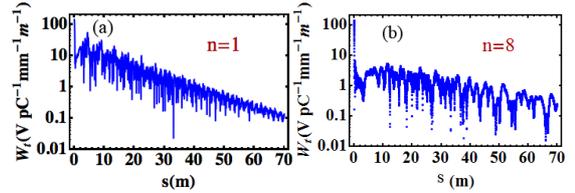

Figure 2: Envelope of the coupled mode wakefield of a single 24 cell structure and of the DDS structure with 8-fold interleaving.

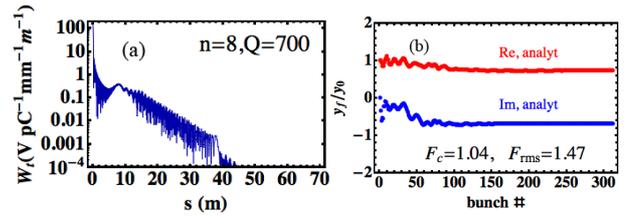

Figure 3: Wakefield (a) for the structure of bandwidth $\Delta\omega/2\pi = 2.1$GHz $= 3.48\sigma_g$ (a) and final normalized amplitudes of the point-like bunches (b).

We have investigated the influence of frequency errors by shifting the bunch spacing by a small fractional amount. This is illustrated in Fig. 4 for both the prescribed Q~700 and for a relaxed Q~1000. We note that even the later case compares well with CLIC_G ($F_{rms}\approx 4.9$) – in this instance, for DDS $F_{rms}$ is below 2.3 for a large range of frequency errors.

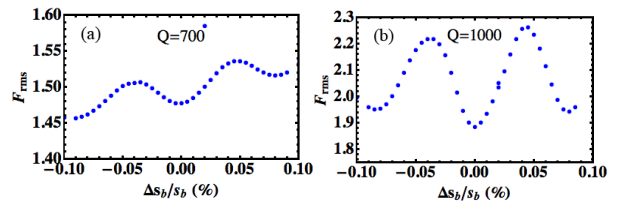

Figure 4: $F_{rms}$ as a function of a systematic bunch spacing error for DDS.

All of these investigations are somewhat idealized, in the sense that neither bunches with realistic bunch length nor energy spread have been included. Additional simulations illustrated in Fig. 5 take these important effects into consideration for both CLIC_G and the DDS structures. Here we also increased the wakefield experienced by the first trailing bunch (in units of $6.6V(pC\cdot m\cdot mm)^{-1}$). It is clear that in all cases, adding these realistic effects helps limit the impact of wakefield on emittance dilution.

Finally we also investigated the potential of completely eliminating the manifolds of the associated higher order mode couplers. This can be achieved by arranging the recoherence position to lay outside the range of the bunch train. Provided the wakefield is well - sampled, then this will occur as the recoherence point proportional to minimum separation of the modes.

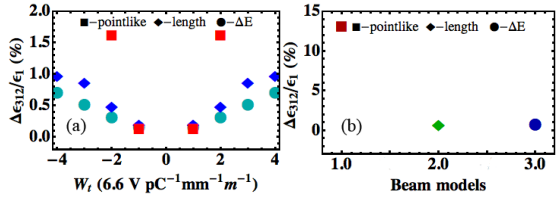

Figure 5: Multi-bunch emittance dilution normalized to the single bunch case as a function of the wake kick on the first trailing bunch for the CLIC_G (a). Different beam models are used: point-like bunches, realistic bunch without initial energy spread (length) and bunches with initial energy spread ($\Delta E$). Also shown is in (b) multi-bunch emittance growth normalized to the single bunch case for DDS (Q=700).

Rather than build a long structure consisting of many cells, we investigated a similar effect - namely interleaving of the modes of successive structures and the impact on emttance dilution of the beam. This is illustrated in Fig. 6. Here we focus on the final offset of the bunch train for a train initially offset by $\Delta y=2\sigma=1\mu m$. Clearly interleaving 30-fold results in a stable bunch train. However, the corresponding minimum mode separation is ~2MHz.

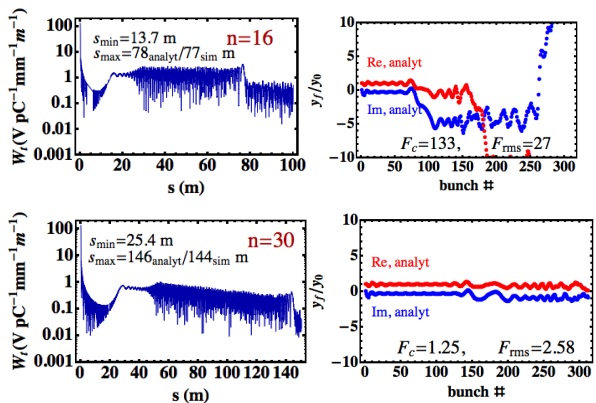

Figure 6: Wakefield for 16 and 30-fold interleaving (left) together with bunch train displacement at the end of the linac (right).

This is a stringent requirement from a manufacturing tolerances perspective – as it is comparable to tolerance imposed on the monopole mode, which is ~1MHz (or ~1μm).

## SUMMARY

This initial beam dynamics study, largely focused on DDS, based on an uncoupled model of the wakefield, indicates a Q~700 is required to maintain the beam quality. In addition the modes must be interleaved with 8 structures. This is summarized in the Table 1.

Finally we note, that pure detuning requires heroically tight manufacturing tolerances – comparable to that imposed on the main accelerating mode. For this reason a structure equipped with strong detuning and manifold suppression is being pursued. The latest design focuses on a cell-to-cell phase advance of $5\pi/6$ [10].

Table 1. Comparison of figures of merit for CLIC

| Design | $F_c$ | $F_{rms}$ |
|---|---|---|
| CLIC_G, Q~10 | 1 | 4.9 |
| DDS×8, Q~2000 | $3\times10^4$ | $2.5\times10^6$ |
| DDS×8, Q=700 | 1 | 1.5 |

## ACKNOWLEDGEMENT

Research leading to these results has received funding from European Commission under the FP7 Research Infrastructure grant no. 227579.

## REFERENCES

[1] H. Braun *et al.*, "CLIC 2008 Parameters", CLIC Note 764.
[2] R.M. Jones, "Wakefield suppression in high gradient linacs for lepton linear colliders", Phys. Rev. ST Accel. Beams 12, 104801 (2009).
[3] A. Grudiev and W. Wuensch, "Design of an x-band accelerating structure for CLIC main linac", LINAC08, 2008.
[4] J. Shi, *et al.*, "Design of a choke-mode damped accelerating structure for the CLIC main linac", IPAC11, 2011.
[5] V.F. Khan, Editorial Series on Accelerator science, Vol.09, ISBN 978-83-7207-951-0.
[6] R.M. Jones, *et al.*, "Analysis and application of microwave radiation from the damping manifolds of the SLAC Damped Detuned structures (DDS)", PAC97, 1997.
[7] A. Latina *et al.*,"Recent improvements in the tracking code PLACET", EPAC 2008.
[8] D. Schulte, "Multi-bunch calculations in the CLIC main linac", PAC09, 2009.
[9] R.M. Jones, "Transverse wakefield suppression using cell detuning assisted by manifold damping for high gradient linear colliders", NIM-A, 657, 2011.
[10] A. D'Elia *et al.*, "Enhanced coupling design of a Detuned Damped structure for CLIC", THPPC022, these proceedings.